\documentstyle[12pt]{article}

\begin{document}
\bibliographystyle{unsrt}
\large
\begin{center}
\underline{Obtaining breathers
in nonlinear Hamiltonian lattices 
} 
\vspace{2cm}\\ \large S. \vspace{0.5cm}Flach$^*$ \\
\normalsize
Max-Planck-Institut f\"ur Physik Komplexer Systeme \\
Bayreuther Str.40 H.16, D-01187 Dresden, Germany \vspace{1cm} 
\\
\end{center}
\normalsize
ABSTRACT \\
We present a numerical method for obtaining high-accuracy
numerical solutions of spatially localized time-periodic
excitations on a nonlinear Hamiltonian lattice.
We compare these results with analytical considerations
of the spatial decay. We show that nonlinear contributions
have to be considered, and obtain very good agreement
between the latter and the numerical
results. We discuss further applications of the method
and results.
\vspace{0.5cm}
\newline
PACS number(s): 03.20.+i ; 63.20.Pw ; 63.20.Ry 
\newline
{\sl Physical Review E, in press} 
\newline
Date: 11/02/94
\\
\\
\\
$^*$ email: flach@idefix.mpipks-dresden.mpg.de
\newpage

\section{Introduction}

The interest in the existence and properties of nonlinear
localized excitations (NLEs) in Hamiltonian lattices
has considerably increased since the work of Takeno et al 
(\cite{st88},\cite{st92} and references therein).
The NLEs are classical solutions and represent localized
vibrations of the constituents (particles) of the Hamiltonian
lattice. 
NLEs are thus similar to breather solutions of the sine-Gordon
equation (a partial differential equation) and sometimes
called discrete breathers.
Remarkably the lattice exhibits discrete translational
invariancy, i.e. no defects or disorder are present. Then the only
reason for the possibility of the existence of NLEs is the
nonlinearity of the system together with its discreteness. The
nonlinearity is needed in order to be able to tune oscillatory
frequencies by changing the energy or amplitude. The discreteness
guarantees a finite band width of the frequency band of
small-amplitude linearized extended excitations (phonons). Consequently
one can avoide resonances between the phonon band and the
frequencies of NLE solutions.

Recently the problem of finding time-periodic NLE solutions
on Hamiltonian lattices was reformulated into finding 
appropriate solutions of a set of coupled algebraic equations
\cite{fw7}.
The variables of this set are the Fourier coefficients of
the NLE solution. This approach opens several possibilities
to obtain analytical and numerical results for NLEs.
Especially spatial decay laws for every Fourier coefficient are
derived, based on a linearization procedure \cite{fw7}. 

In the present work we will first formulate a new mapping
in order to numerically solve the mentioned set of algebraic equations.
We obtain high-precision numerical solutions for these
equations. The level of accuracy allows us to check
analytical predictions. Next we reconsider the spatial
decay problem of the NLE solution and find, that under
certain conditions the previously applied linearization procedure
does not hold without exceptions. 
We obtain nonlinear corrections and
formulate conditions for their restricted applicability.
Finally we make use of the availability of high-precision numerical
solutions and test our analytical predictions.

\section{Formulation of the problem}

We consider a system of classical interacting particles
arranged on a certain lattice. By that we mean that the
positions of all particles in the groundstate
(potential minimum) define a regular lattice structure.
The potential energy of the system, expressed in
the relative displacements of the particles from
their groundstate positions exhibits a minimum
if all displacements are zero.
Since we are interested in the easiest demonstration
of what we might call a general phenomenon, we restrict
ourselfes to the case of
a one-dimensional lattice with one degree of freedom per unit cell.
Also we choose nearest neighbour interaction.
We show later that these restrictions are at best
of some technical use, i.e. more complicated
systems will exhibit the same behaviour.
The Hamilton function of our system can be then given in the form
\begin{equation}
H = \sum_{l}\left[ \frac{1}{2}P_{l}^2
+ V(X_{l})  +
\Phi(X_l - X_{l-1}) \right ] . \label{1}   
\end{equation}
Here $P$ and $X$ denote canonically conjugated momentum
and displacement of a particle and the label $l$ counts the
unit cells.
The on-site potential $V(z)$ and the interaction potential
$\Phi(z)$ can be choosen in a Taylor expansion representation:
\begin{eqnarray}
V(z)=\sum_{\alpha=2,3,...}\frac{v_{\alpha}}{\alpha !}z^{\alpha}
\;\;, \label{2-1} \\
\Phi(z)=\sum_{\alpha=2,3,...}\frac{\phi_{\alpha}}{\alpha !}
z^{\alpha}\;\;. \label{2-2}    
\end{eqnarray}
Finally the equations of motion are given by
\begin{equation}
\ddot{X}_l = - \frac{\partial H}{\partial X_l} \;\;. \label{3}      
\end{equation}
We search for a solution of (\ref{3}) in the form
\begin{equation}
X_l(t)=X_l(t+T_1)\;\;,\;\;X_{l \rightarrow \pm \infty} 
\rightarrow 0 \;\;. \label{4}    
\end{equation}
If solution (\ref{4}) exists, we can introduce a Fourier series
representation for $X_l(t)$:
\begin{equation}
X_l(t)=\sum_{k = -\infty}^{+\infty} A_{kl} {\rm e}^{ik\omega_1 t}
\;\;, \label{5}   
\end{equation}
where $\omega_1=2\pi / T_1$. Let us separate the linear terms
in the displacements $X$ on the right hand side of 
(\ref{3}) from the nonlinear ones:
\begin{equation}
\ddot{X}_l = -v_2X_l - \phi_2 (2X_l - X_{l-1} - X_{l+1})
+ F^{nl}_l ({X_{l'}}) \;\;. \label{6}      
\end{equation}
The nonlinear part $F^{(nl)}_l$ of the force contains only
higher than linear terms in the displacements: 
\begin{equation}
F^{(nl)}_l = -\sum_{\alpha=3,4,...}\left[
\frac{v_{\alpha}}{(\alpha-1) !}X_l^{\alpha -1} + 
\frac{\phi_{\alpha}}{(\alpha - 1) !} ((X_l - X_{l-1})^{\alpha -1}
- (X_{l+1} - X_l)^{\alpha - 1} ) \right] \;\;. \label{7}    
\end{equation}
Since all functions
$X_l(t)$ of our proposed solution (\ref{4}) are periodic with the
same period $T_1$, we can also expand $F^{(nl)}_l$ into a Fourier
series:
\begin{equation}
F^{(nl)}_l (t) = \sum_{k = - \infty}^{+ \infty}
F^{(nl)}_{kl} {\rm e}^{ik\omega_1 t} \;\;. \label{8}    
\end{equation}
The Fourier coefficients $F^{(nl)}_{kl}$ will be some functions
of the coefficients $A_{k'l'}$ from (\ref{5}) (depending on the
given form of the potential functions (\ref{2-1}),(\ref{2-2})).
Inserting (\ref{5}) into (\ref{3}) and using (\ref{8}) we finally obtain:
\begin{equation}
k^2 \omega_1^2 A_{kl} = v_2 A_{kl} + \phi_2 (2A_{kl} - A_{k,l-1}
- A_{k,l+1}) + F^{(nl)}_{kl}\;\;. \label{9}      
\end{equation}
The infinite set of equations (\ref{9}) (for all $k$) represents
a coupled set of equations for the Fourier coefficients $A_{kl}$.

This set of equations has to be analyzed in order to obtain
nonlinear localized excitations, i.e. (cf. (\ref{4})) solutions
with
\begin{equation}
A_{k,l \rightarrow \pm \infty} \rightarrow 0 \;\;. \label{10}     
\end{equation}
Let us shortly review what is known about the solutions of
(\ref{9}) with condition (\ref{10}). In \cite{fw9} it was shown, that
the finding of a nonlinear localized excitation is equivalent
to obtaining common points (homoclinic points) 
of two separatrix manifolds of
a certain mapping. It was also shown there, that
if a solution is can be found for a given system, then it is
structurally stable with respect to perturbations of the
Hamilton function and thus of generic type. Also in \cite{fw9}
an existence proof of NLEs was given for a system with $V(z)=0$
and $\Phi(z)=(1/2m)z^{2m}$ ($m=2,3,...$).
Furthermore MacKay and Aubry \cite{ma94}
have shown, that in the limit of weak interaction $\Phi(z)$
and any anharmonic potential $V(z)$ NLEs exist, remarkably for
any lattice dimension, and also for larger interaction ranges
and numbers of degrees of freedom per unit cell. Consequently
the above considered simplest case is not different from
more complicated (and therefore more realistic) models.
A systematic study of one-dimensional \cite{fw2},\cite{fw6} 
and two-dimensional
\cite{fw8} lattices shows that there is absolutely no difference
in the appearance and properties of NLEs.

\section{The method of finding solutions - designing a map }

Let us reformulate Eq. (\ref{9}) into a map in order to
obtain a numerical procedure for finding NLE solutions.
For that we rewrite (\ref{9}) in two ways:
\begin{eqnarray}
A^{(i+1)}_{kl} = \frac{1}{k^2\omega_1^2} \left[ (v_2 + 2\phi_2)A^{(i)}_{kl}
- \phi_2 (A^{(i)}_{k,l-1} + A^{(i)}_{k,l+1}) + F^{(nl)}_{kl}
({A^{(i)}_{k'l'}}) \right] \;\;, 
\label{11} \\
A^{(i+1)}_{kl} = \frac{1}{v_2} \left[ (k^2\omega_1^2 -2\phi_2)A^{(i)}_{kl}
+ \phi_2(A^{(i)}_{k,l-1} + A^{(i)}_{k,l+1}) - F^{(nl)}_{kl}(
{A^{(i)}_{k'l'}}) \right] \;\;.
\label{12}      
\end{eqnarray}
Here the notion of the upper labels $(i+1)$ and $(i)$ can be ignored
for a moment. For clarity we explicitely indicate in (\ref{11}),(\ref{12})
the dependence of the coefficients $F^{(nl)}_{kl}$ on the coefficients
$A_{kl}$. Both equations (\ref{11}),(\ref{12}) are equivalent to
each other with respect to solutions. If we now take into account
the upper labels $(i+1)$ and $(i)$, then these two equations
in fact become two maps. Given the coefficients $A_{kl}$ at
the $i$-th step, we can calculate the coefficients $A_{kl}$
at the $(i+1)$st step. If we find a fixed point of any of the two
maps, then it will be also a fixed point of the second map.
Each fixed point of these maps is a solution of the original
equation (\ref{9}).

One fixed point can be immediately found - it is $A_{kl}=0$
for all $k,l$. If we linearize the map around this fixed point
and assume for a moment $\phi_2=0$ we obtain the eigenvalues
\begin{eqnarray}
\lambda^{(\ref{11})}_{kl} = \frac{v_2}{k^2\omega_1^2} \;\;, \label{13} \\
\lambda^{(\ref{12})}_{kl} = \frac{k^2 \omega_1^2}{v_2} = \frac{1}{\lambda
^{(\ref{11})}}\;\;. \label{14}     
\end{eqnarray}
Here the upper labels of the eigenvalues indicate the
equation number of the map they
belong to. 
If now $\phi_2$ is smoothly increased from zero, then
the eigenvalues (\ref{13}),(\ref{14}) will also smoothly change \cite{jhw92}.
Consequently we might expect some control over the values of the
eigenvalues in the case of nonzero interaction.
As for the eigenvectors, any orthonormal set of basis vectors
would be an eigenvector base for the linearized maps in the
case $\phi_2=0$. For clarity we can always choose the eigenvector
set which appears in the limit $\phi_2 \rightarrow 0$. Then the 
eigenvectors are also smoothly changed by increasing $\phi_2$
from zero.   

As it follows from (\ref{14}), both eigenvalues for a given
pair of $(k,l)$ are positive and inverse to each other. Consequently
one of the eigenvalues will be larger than one and the second
eigenvalue will be smaller than one. In order to find a NLE solution
we design a new map out of the two maps (\ref{11}),(\ref{12}).
This new map can be composed out of the two maps (\ref{11}),(\ref{12})
in an appropriate way depending on the given problem one wants
to solve. Let us explain the strategy of this designing. Suppose
we want to find a NLE solution with a given frequency $\omega_1$,
which is centered on a lattice site $l=0$.
Then we choose out of the two maps
(\ref{11}),(\ref{12}) the one which yields  
an eigenvalue larger than 1 for the pair
$(k = \pm 1, l=0)$ according to (\ref{11}),(\ref{12}). For all other
pairs $(k,l)$ we choose the map which yields an eigenvalue smaller than 1
(note that this choice might be still different for different $k$).
By that we achieve the following. If we choose as an initial condition
for our designed map a set of the $A_{kl}$ with $A_{k=\pm 1,l=0}=\delta$,
$A_{k\neq \pm 1, l \neq 0}=0$, then we can expect that starting iterating
our map we essentially get growth in the direction of the unstable
eigenmode at $(k=\pm 1,l=0)$, whereas all other eingenmodes are not
excited (eigenvalues lower than one). Because we have designed a map
in which the translational invariancy with respect to $l$ is broken
at $l=0$, we also know that the corresponding eigenvector is of local
character for small $\phi_2$. A way to estimate whether this statement
is correct is to calculate the ratio of the off-diagonal terms
in the linearized map (which are roughly given by $\phi_2 / v_2$)
over the difference between the diagonal terms:
\begin{equation}
\frac{\phi_2}{v_2} (\frac{v_2}{\omega_1^2} - \frac{\omega_1^2}{v_2})
= \frac{\phi_2 \omega_1^2}{v_2^2 - \omega_1^4}\;\;. \label{15}    
\end{equation}
This ratio has to be small enough in order to be sure that the 
above given arguments hold. Once our iteration yields local growth
for the Fourier coefficients, we can hope that the nonlinear terms
of the map, which become important for large enough $A_{kl}$, will
lead the iteration to the fixed point which corresponds to an NLE solution.
There is absolutely no guarantee for that to happen. We have no 
knowledge about the properties of this fixed point, even if we would
know for sure that the NLE solution exists (which means in turn
that the fixed point exists). All the designing is focussing on
is to get the right initial growth (namely local growth) in our iteration.

If we would search for a NLE which is centered between two lattice
sites $l=0,1$, all we would have to change is to add an unstable
map for the pair $k=\pm 1, l=1$, and depending on the symmetry
of the solution (in-phase or out-of-phase motion) to choose equal
or opposite signs for the initial perturbations in $A_{k=\pm 1,l=0}$ and
$A_{k=\pm 1, l=1}$.

\section{Solutions}

We implemented two numerical realizations for the designed map.
The main difficulty is to account for the nonlinear terms
$F^{(nl)}_{kl}$ (\ref{8}). In this paper we report on results for
systems with $V(z)$ and $\Phi(z)$ being finite order polynomials
in $z$. Then one way of calculating the nonlinear terms
is to evaluate all possible combinations of products of
the Fourier coefficients $A_{kl}$. 
We will call this method {\sl polynomial approach}.
As an example we give the
contribution of the first term in the sum on the right-hand side
of equation (\ref{7}):
\begin{equation}
 F^{(nl)}_{kl} = \sum _{\alpha=3,4,...}\frac{v_{\alpha}}{(\alpha - 1)!}
\sum_{k_1,k_2,...,k_{\alpha - 1} = - \infty}^{+ \infty}
A_{k_1l}A_{k_2l} ... A_{k_{\alpha - 1}l} \delta_{k,(k_1 + k_2 + ...
+ k_{\alpha - 1})} \;\;. \label{4-1}  
\end{equation}
Here $\delta_{a,b}$ stand for the Kronecker symbol with integer $a,b$.

The polynomial approach can become very unefficient if the number and
order of nonlinear terms in (\ref{4-1}) become large (in fact
$\alpha=5$ can already pose serious computing problems).
If the potential functions are not given by finite order polynomials
then the polynomial approach breaks down. In this case it is more useful
to numerically integrate the function $F_l^{(nl)}(t)$:
\begin{equation}
 F^{(nl)}_{kl} = \frac{1}{T_1} \int_{-T/2}^{T_2} F_l^{(nl)}(t)
{\rm e}^{-ik\omega_1 t}{\rm d}t \;\;. \label{4-2}    
\end{equation}
We will call this approach {\sl Fourier integral approach}.
Here $T_1 = 2 \pi / \omega_1$ is the period of the NLE solution.
If we know $A_{kl}$ for all $(k,l)$, then we can immediately
calculate (\ref{4-2}) using (\ref{7}).
In order to test the accuracy of our calculations we can use
both approaches and thus compare the two results, which have to
be equivalent.

Another important restriction on the numerical approach is
the finiteness of the $k$-sums. We have to introduce by hands
a cutoff in $k$-space: $ A_{kl} = 0$ for $|k| \geq k_{max}$. 
This cutoff will be justified only
if the numerically evaluated Fourier coefficients $A_{kl}$ 
will drop to small enough values at the cutoff $k_{max}$. 
Here small enough will be defined through the accuracy of the
numerical calculations. We will stop the iteration if the
sum 
\begin{equation}
 \sum_{k,l}|A_{kl}^{(i)} - A_{kl}^{(i-1)}| < 10^{-10}\;\;. \label{error}    
\end{equation}

In this paper we will study three models. Model I is characterized by
$V(z)= 1/2z^2 - 1/4 z^4$ and $\Phi(z)=1/2 C z^2$. Model II has
the potential functions $V(z)=z^2 - z^3 + 1/4 z^4$, $\Phi(z)= 1/2 C z^2$.
Model III is defined through $V(z)=1/2 z^2 + 1/4  z^4$,
$\Phi(z)=1/2 C z^2$ . Model I is characterized by
a local minimum for $z=0$ and a global instability for $|z| \geq 1$.
Here we will restrict ourselfes to small enough amplitudes such that
we avoid the instability region. Model II is the well known $\Phi^4$-lattice
model (after a change of variables $z'=z-1$ one arrives at the
symmetric form of the double-well onsite potential $V(z')=
1/4 (z'^2 - 1)^2$). This model has in fact two identical groundstates
$z'=\pm 1$ or $z=0,2$. Without loss of generality we have choosen
$z'=-1$ as the groundstate around which we expand the potential functions.
Model III differs from Model I by the sign in front of the quartic term
in $V(z)$. Consequently there is no instability in this model.
In all models the parameter $C=\phi_2$ regulates the bandwidth
of phonons obtained after a linearization of the
equations of motion around the groundstate:
\begin{equation}
v_2 \leq \omega_q^2 \leq v_2 + 4C \;\;. \label{4-3}      
\end{equation}
Here $\omega_q$ is the frequency of a phonon wave with wave number $q$.
For our models we have: model I - $v_2=1$; model II - 
$v_2=2$; model III - $v_2=1$. In our calculations we will use $C=0.1$.

In all three model cases the phonon band has a nonzero lower
band edge $v_2$. In order to solve the map for the Fourier coefficients
we have to know (or be able to predict) whether the NLE frequency
can be below or above the phonon band (or may be below and above).
There exists a rather straightforward method to do so. 
First we have to know the symmetry of the solutions with respect
to their position on the lattice. In most cases (even supported by
rigorous proofs for some cases) we can expect two solutions:
a NLE centered on a lattice site (meaning that the
center of energy of the energy distribution
of the solution has a maximum on a lattice site) which we call XC solution;
and a NLE centered between two lattice sites which we call XN solution.
(this notation was first used in \cite{cp90}). Now all one has to do
is to consider two possibilities: that the NLE frequency might be
below or above the phonon band. As it was shown in \cite{fw7}, this
option defines whether the motion of particles in the solution
is of in- or out-of-phase character. In \cite{fw2},\cite{fw8}
it was explained how to construct a certain effective potential
out of these input informations. The motion of a particle in this
effective potential can be then analyzed with respect to the
energy dependence of its oscillation frequency. This energy dependence
is qualitatively (and often even quantitatively) the same as the
energy dependence of the NLE. If e.g. the initial assumption was
'frequency below the phonon band', but the effective potential 
yields only frequencies above the phonon band, then the initial
assumption was wrong (and vice versa). Let us give the form of
the effective potentials for the 4 different cases:
\begin{eqnarray}
 V_{eff}(z) = V(z)+ \Phi(z)+\Phi(-z) \;\;,\;\;XC\;\;,\;\;\omega_1 < \omega_q
\;\;, \label{4-4} \\
V_{eff}(z) = 2V(z) + \Phi(z) + \Phi(-z)\;\;,\;\;XN\;\;,\;\;\omega_1 < \omega_q
\;\;, \label{4-5} \\
V_{eff}(z) = V(z) + \Phi(z) + \Phi(-z) \;\;,\;\;XC\;\;,\;\;,\;\;
\omega_1 > \omega_q
\;\;, \label{4-6} \\
V_{eff}(z) = V(z) + V(-z) + \Phi(2z) + \Phi(z) + \Phi(-z)
\;\;,\;\;XN\;\;,\;\;\omega_1 > \omega_q \;\;. \label{4-7}        
\end{eqnarray}
As it follows from the analysis, model I allows only for cases 
(\ref{4-4}),(\ref{4-5}) ($\omega_1 < \omega_q)$. For model III we
find the reverse case $\omega_1 > \omega_q$. For model II and
small energies we find $\omega_1 < \omega_q$, whereas for large
energies also $\omega_1 > \omega_q$ can be realized.

Having this information, we can proceed with the numerical treatement
of the map for the Fourier coefficients. 
In Fig.1 we show an
XC NLE solution
as obtained for model I and the frequency $\omega_1=0.8$. 
We used periodic boundary conditions (PBC), 100 lattice sites and
$k_{max}=30$. Because of the symmetry of the potential functions all
Fourier coefficients with  even $k$ vanish. In our calculations
(both with the polynomial and Fourier integral approach) we however
did not use this information for the calculation. Both methods
yield exactly the same solution, in particular the even Fourier
coefficients indeed vanish. In Fig.1 the dependence of $|A_{kl}|$
on the lattice site $l$ is shown in a semi-logarithmic plot.
The actual data are given as open squares. Data for equal values
of $k$ are connected with solid lines. The value of $k$ increases
from top to bottom: $k=1,3,5,...$. As is clearly seen in Fig.1,
Fourier coefficients for each $k$ show up with a logarithmic
decay in $l$. What is not observable in the plot of Fig.1 is that
the Fourier coefficents for a given $k$ alternate their signs
as one increases or decreases $l$ by 1. Only for $k=1$ the sign
of all $A_{1l}$ is the same, independent on $l$. The Fourier 
coefficients for negative $k$ are identical with the corresponding
coefficents for positive $k$: $A_{kl}= A_{-k,l}$. Let us also 
mention, that we have checked any finite size effects. We have
computed the same solution for 30 lattice sites, and found
{\sl practically no} difference in the values of the Fourier
coefficients. This appears to be logical, since the solution
we find is very localized, so that it does not matter, how large
the system is. Consequently we can be sure that we find a solution
which is also a solution of the infinite problem.

In Fig.2 we show an XN NLE solution for the same model I, where
all other parameters are as in case of the solution in Fig.1.
All the comments from Fig.1 apply to Fig.2. 
 
Let us consider model II. We choose a frequency $\omega_1=1.3$
which is below the phonon band. Because of the assymetry of
the corresponding onsite potential $V(z)$ Fourier components
with even $k$ will be nonzero. The result for an XC NLE is shown
in Fig.3. Data belonging to one $k$-value are again connected
with solid lines. The value for k from top to bottom is
$k=1,0,2,3,4,...$. The signs of $A_{kl}$ are positive for
$k=0,1,3$, negative for $k=2$ and alternating for all other $k$-values.
We also created a similar XN NLE solution with same frequency for
model II which behaves similar to the XC solution in Fig.3.

In Fig.4 we show the result for an XC NLE solution for model III.
Here the frequency $\omega_1=1.3$ is choosen to be above the
phonon band. As in the case of 
model I the symmetry of the potential functions
causes all even $k$ Fourier coefficients to vanish. The $k$-values
for the nonzero odd $k$ Fourier coefficients in Fig.4 increase
from top to bottom : $k=1,3,5,...$. For all odd $k$-values the
signs of the Fourier coefficients alternate as we change the
lattice site $l$ by 1. Again we also created a similar XN NLE solution.

\section{The spatial decay revisited}

In \cite{fw7} the spatial decay of the Fourier components
$A_{kl}$ was studied. The main idea was to start with (\ref{9})
and rewrite it in the form
\begin{equation}
 A_{k,l+1} = \frac{v_2}{\phi_2} A_{kl} + 2A_{kl} - A_{k.l-1}
- \frac{k^2\omega_1^2}{\phi_2} A_{kl} + \frac{1}{\phi_2} F_{kl}^{(nl)}
\;\; . \label{5-1}   
\end{equation}
Equation (\ref{5-1}) can be interpreted as a map: knowing all Fourier
coefficients at lattice sites $l$ and $(l-1)$ one can calculate
the coefficients at $(l+1)$. Naturally one can iterate also
to negative values of $l$ (i.e. one just exchanges $(l+1)$ with
$(l-1)$ in (\ref{5-1})). If we iterate into the tails of a NLE solution,
then because of (\ref{10}) it was suggested in \cite{fw7} that the
nonlinear contributions $F_{kl}^{(nl)}$ which are at least of
second order in $A_{k'l'}$ can be neglected. Then equation (\ref{5-1})
is transformed into an infinite set of two-dimensional maps for
each value of $k$. As a result exponential decay laws are obtained 
\cite{fw7}:
\begin{equation}
A_{kl} \sim \left( {\rm sgn}(\lambda_k)\right) ^l {\rm e}^{{\rm ln}
|\lambda_k | l}\;\;, \;\; \lambda_k = 1+\frac{\kappa_k(\omega_1)}{2}
\pm \sqrt{(1 + \frac{\kappa_k(\omega_1)}{2})^2 - 1}\;\; \label{5-2}    
\end{equation}
with $\kappa_k(\omega_1) = (v_2- k_1^2 \omega_1^2)/(\phi_2)$.
The sign in (\ref{5-2}) hase to be choosen in order to obey the
condition $|\lambda_k| \leq 1$. One of the interesting properties
of (\ref{5-2}) is that all one has to know about the problem
is the frequency of the NLE solution and the positions of
the phonon band which are given by ($v_2,\phi_2$). 
Consequently several predictions
were drawn in \cite{fw7} on this basis which were successfully 
checked. 

However in order to justify the dropping of the nonlinear contributions
in (\ref{5-1}) one has to insert the predicted decay laws as obtained
from the linearization into the full equations 
and show that the nonlinear terms are indeed
negligible.  
Let us consider the contributions
of the nonlinear terms as given in (\ref{4-1}).
First we note that the function $d(k) = {\rm ln}(|\lambda_k|)$ 
is by definition always negative. $d(k)$ is defined on a discrete
lattice (integer $k$) and $d(k)=d(-k)$. Let us consider $k \geq 0$.
Then $d(k)$ has a single maximum either for one value of $k=k_m$
or for two subsequent values $k=k_m$ and $k=(k_m + 1)$. For $k \geq k_m$
the function $d(k)$ is monotonically decreasing with increasing $k$.
Then the linearization in (\ref{5-1}) is violated for a given $k_0$
if there exists a sequence of $k_1,k_2,...,k_{\alpha-1}$ such that
\begin{equation}
d(k_0) \leq (d(k_1) + d(k_2) + ... + d(k_{\alpha - 1}))  \label{5-3}      
\end{equation}
holds together with the condition (see (\ref{4-1}))
\begin{equation}
k_0 = \pm k_1 \pm k_2 \pm ... \pm k_{\alpha - 1}\;\; . \label{5-4}    
\end{equation}
The reason is that in this case a given nonlinear combination
is decaying in space weaker than the corresponding linear term.

Since equations (\ref{5-3}),(\ref{5-4}) can never be satisfied
for $k_0 = k_m$ it follows that the linearization is always
justified for $k_m$. Let us consider $k > k_m$. If the
sequence of $(k_1,k_2,...,k_{\alpha-1})$ yields violation of the
linearization, then all $k$-numbers from this sequence have to
be of lower value than $k$. Increasing $k$ step by step starting
from $k_m$ we can then account for all possible nonlinear corrections
of the result of linearization. Since $d(k) \sim -2{\rm ln}(k)$ for
large values of $k$, it follows from a numerical analysis 
that for each value of $\alpha$
there will be a $k_+$ such that no nonlinear corrections appear
for $k > k_+$. Also for a given function $g(k)$ there will be 
a $\alpha_+$ such that no nonlinear corrections appear for any value
of $k$.

Let us consider the numerical results for NLEs from the previous
section in the light of the 
above given considerations. The case shown in Fig.1 was analyzed
in \cite{fw7}. Fig.5 in \cite{fw7} shows very good agreement between
the $k$-dependent exponents of the decay as found from
the numerical solution (Fig.1) and the result of the linearization.
A careful check of the nonlinear corrections indeed show that
the linearization result holds without exception. 
The same result is valid for
the XN NLE in Fig.2. 

The XC NLE solution shown in Fig.3 can not be described in
its decay properties by the linearization result only. In Fig.5
we show the exponents as calculated from Fig.3 (squares) and
the predicted exponents from the linearization procedure
(filled circles). Clearly
for $k=2$ (and also less visible for $k=3$) the numbers differ.
The exponents for the solution are 
\[
\begin{array}{ccc}
k & { \rm num. result} & {\rm linearization} \\
0 & -1.3202 & -1.3415 \\
1 & -0.6904 & -0.6898 \\
2 & -1.3796 & -1.6588 \\
3 & -2.0748 & -2.1143 \\
4 & -2.3957 & -2.3951 \\
5 & -2.6018 & -2.6026 \\
6 & -2.7663 & -2.7682 \\
\vdots & \vdots & \vdots    
\end{array}
\]
Nonlinear corrections can appear from terms with $\alpha=3,4$.
For $\alpha=3$ we can find the following corrections.
For $k=2$ the nonlinear contribution $A_{1l} A_{1l}$ yields
an exponent $(-0.6904 - 0.6904) = -1.3796$, which is exactly what is found from
the numerical result and is larger than the value of
the linearization result $-1.6588$. 
The sign of $A_{2l}$ should alternate
according to the linearization result. 
If the nonlinear terms in (\ref{5-1}) dominate, the sign
of $A_{2l} \sim -A_{1l}^2$ should be negative - exactly as
found from the numerical result (cf. previous section).
For $k=3$ the nonlinear contribution $A_{1l}A_{2l}$ 
(with the corrected exponent for $A_{2l}$) yields
an exponent $(-0.6904 - 1.3796) = -2.0700$ which is again very close to the
numerical result and larger than the linearization result $-2.1143$.
The sign of $A_{3l}$ should alternate according to the linearization
result. If the nonlinear terms in (\ref{5-1}) dominate the sign
of $A_{3l} \sim -A_{1l}A_{2l}$ should be positive ($A_{1l} > 0$
is found from the numerical solution) - exactly as found from
the numerical result (cf. previous section).
The only nonlinear contribution from the nonlinear term
with $\alpha = 4$ appears for $k=3$ ($A_{1l}^3$) and is equivalent to
the $\alpha=3$ contribution ($A_{1l}A_{2l}=A_{1l}A_{1l}^2$) with respect to
the value of the exponent as well as with respect to
the sign of the correction. 
For $k > 3$ no
nonlinear corrections are found - consequently the linearization
result coincides with the numerical findings both in the
value of the exponent and in the sign of the Fourier components
(alternating).

The XC NLE solution shown in Fig.4 has also nonlinear corrections
in the spatial decay.  In Fig.6 we show the exponents as calculated
from Fig.4 (squares) and the predicted exponents from the
linearization procedure (filled circles). For $k=3$ the results are
different. The exponents for the solution are
\[
\begin{array}{ccc}
k & {\rm num. result} & {\rm linearization} \\
1 & -0.6722 & -0.6709 \\
3 & -1.9910 & -2.1464 \\
5 & -2.6103 & -2.6133 \\
7 & -2.9114 & -2.9117 \\
9 & -3.1324 & -3.1325 \\
\vdots & \vdots & \vdots
\end{array}
\]
Only nonlinear terms with $\alpha = 4$ are present. For $k=3$
the nonlinear contribution $A_{1l}A_{1l}A_{1l}$ yields an exponent
$(-0.6722 -0.6722-0.6722)=-2.0166$ - very close to the
numerical result and larger than the value of the linearization
result $2.1464$. Since the sign of $A_{1l}$ alternates, so does
the sign of $A_{3l} \sim A_{1l}^3$. For all other $k \neq 3$
the nonlinear terms give no corrections to the linearization result.

Nonlinear corrections can be expected if one of the multiples
of the frequency $\tilde{k}\omega_1$ comes close to the phonon band -
then the corresponding spatial decay exponent comes close to
zero and can easily contribute to the decay properties of
Fourier coefficients with e.g. $k= (\alpha - 1)\tilde{k}$.

\section{The decay in $k$-space}

So far we have considered the spatial decay of NLE solutions.
However as it was already explained, in order to obtain
numerical solutions we have to introduce a cutoff in
$k$-space. This cutoff will be only justified if the 
Fourier components decay in the $k$-space. From the examples
in Figs.1-4 this decay is clearly visible. In this section we will
obtain some analytical results for the decay in $k$-space
for a solution where no nonlinear corrections to the linearization
result (of the spatial decay, see previous section) appear.

As it appears from (\ref{5-2}) the $k$-dependence of
the Fourier coefficients is given
by 
\begin{equation}
A_{kl} \sim s(k) {\rm e}^{-{\rm ln}(|\lambda_k|)|l|} \label{6-1}   
\end{equation}
where the function $s(k)$ describes the $k$-dependence of
the Fourier coefficients at lattice site $l=0$ (for convenience
we assume that the NLE is created with center at $l=0$; 
in case of reference to numerical results a shift in the
lattice site numbers is implied). If we consider the XC NLE solution
in Fig.1, then we observe that ${\ln}(s(k)) \sim -k $. For
large values of $k$ 
\[
\lambda \approx - \frac{\omega_1^2 k^2}{\phi_2}
\]
and consequently
\begin{equation}
|A_{kl}| \sim k^{-2|l|} s(k) \;\;. \label{6-2}    
\end{equation}
In case of Fig.1 this implies that the decay of the Fourier components
in $k$-space is given by a product of an exponential decay and
a power law decay, where the power law decay is lattice site dependent -
the farer from the center of the NLE, the larger the exponent of
the power law decay. In fact according to (\ref{6-2}) the exponent
of the power law decay should increase by $2$ as one switches from
a lattice site to its nearest neighbour. 

In Fig.7 we plot the normalized Fourier components $|A_{kl}|/s(k)$ 
from Fig.1 as a function of $k$ in a log-log plot.Data with same
values of $l$ are connected with lines. Clearly power laws are observed.
In Fig.8 the $l$-dependence of the power law exponent
is shown by plotting the differences between power law exponents
on neighbouring lattice sites - as predicted this difference appears
to be approximately $-2$ (deviations for larger values of $l$
appear because of the smallness of the unrenormalized Fourier
coefficients).

From the results in this section we can follow that the decay
of the Fourier components in $k$-space is as faster as the
distance from the NLE center increases.

\section{A variational approach}

So far we have implemented a numerical method of obtaining
NLE solutions and have used the numerical results
in order to verify and improve several analytical findings. 
For completeness we show in this section that NLE solutions
can be obtained out of a variational problem, i.e. we
can define a function of the Fourier coefficients such
that a NLE solution will correspond to an extremum of
this function. This result can be useful both in considering
an independent method of finding numerical NLE solutions as
well as in performing existence proofs for NLEs in lattices
with dimensions higher than one and in considering the
analogous properties of quantum lattices.

Let us consider the action $S$ of a given classical lattice:
\begin{equation}
S = \int_{0}^{t}{\rm d}\tau L  \;\; \label{7-1}   
\end{equation}
where the Lagrange function $L$ is given by
\begin{equation}
L = \sum_l \dot{X}_l^2 - H \;\;, \label{7-2}      
\end{equation}
and the Hamilton function $H$ is given in (\ref{1}). We search
for a periodic solution (closed trajectory) with period 
$t=2\pi / \omega_1$.
Inserting (\ref{5}) into (\ref{7-1}) and integrating
over the period we find for the action $S_p$ of a periodic solution
\begin{eqnarray}
S_p\frac{\omega_1}{2\pi} = \sum_l \sum_{k}\frac{1}{2}k^2\omega_1^2 
A_{kl}A_{-k,l}
\nonumber \\
- \sum_l \sum_{\alpha =2}^{\infty}v_{\alpha}\frac{1}{\alpha !}
\sum_{k_1,k_2,...,k_{\alpha}}A_{k_1l}A_{k_2l}...A_{k_{\alpha}l}
\delta_{0,(k_1+k_2+...+k_{\alpha})} - ... \label{7-3}   
\end{eqnarray} 
Here the $...$ in (\ref{7-3}) implies the evaluation of the
interaction terms (\ref{3}), which look similarly to the term
in the second row of Eq. (\ref{7-3}) except that the lattice site labels
in one product can come from two neighbouring lattice sites.

Let us consider the derivative of $S_p$ with respect to a given
$A_{kl}$:
\begin{equation}
\frac{\omega_1}{2\pi}\frac{\partial S_p}{\partial A_{kl}} =
 k_2\omega_1^2 A_{-k,l} - 
\sum_{\alpha=2}^{\infty} v_{\alpha}
\frac{1}{(\alpha - 1)!}
\sum_{k_1,k_2,...,k_{\alpha - 1}} 
A_{k_1l}A_{k_2l}... A_{k_{\alpha -1 }l} \delta_{0,(k + k_1 + k_2 + ...
+ k_{\alpha - 1})} - ... \;\;. \label{7-4} 
\end{equation}
Now we observe that if a periodic solution on the lattice is found,
i.e. if equation (\ref{9}) holds, then for these solutions 
the right hand side of (\ref{7-4}) vanishes (if one replaces
$k$ by $-k$ in (\ref{9})). Consequently the function $S_p$
from (\ref{7-3}) has to have an extremum with respect to
variations of the Fourier coefficients $A_{kl}$ for a periodic
solution. Also the inverse is true: if $S_p$ has an extremum
with respect to variations of the Fourier coefficients
\begin{equation}
\frac{\partial S_p}{\partial A_{kl}} = 0 \label{7-5}
\end{equation}
for all $(kl)$ then the set of Fourier coefficients which corresponds
to this extremum is a periodic solution of the equations of motion.

\section{Discussion}

We have shown the construction of a map for the
Fourier coefficients of a periodic solution on a
lattice. This map is specifically designed in order
to obtain spatially localized time-periodic solutions.
The usefulness of this method is first that very little
(in fact nothing specific) has to be known about the
solution one searches for, i.e. the initial condition
(starting distribution of the Fourier coefficients) has
to reflect only the localized character of the solution one
looks for. This circumstance makes our approach much
more easy than e.g. the approach used in \cite{cp90}. There
also a map was designed in order to obtain localized time-periodic
solutions. However the initial condition had to be choosen
to be very close to the final result. In other words one had
to obtain information about the final solution before even 
starting the map. 

Secondly with our method we derive final 
solutions with high numerical precision. This precision allows
for the numerical analysis of several problems, e.g. the
dynamical stability of the derived solutions, the 
study of scattering events of plane waves by the localized
solutions and others. Let us explain this using the
scattering problem. In \cite{fwdna} the scattering (reflection)
of phonon waves by a single localized solution was studied
for a one-dimensional lattice. The maximum amplitude for
the localized solution was of the order of 1, and
the errors of the amplitudes for the same solution were of the
order of $10^{-3}$ leading to uncertainties in the energy
distribution of order $10^{-6}$. The infalling phonon wave had
to have an energy density of the order of $10^{-4}$ in order
to ensure nearly linear dispersion. The transmittion of phonons
was increasingly suppressed with increasing wave number.
Since the squared transmittion coefficient was observable only
for values down to $10^{-2}-10^{-3}$ the maximum wave number
under study was around $0.4 q_{ZB}$ (here $q_{ZB}$ means the
zone boundary wave number). With the accuracy of the final solutions
as derived in this work ($10^{-15}$ in amplitudes, $10^{-30}$
in the energy density) one should be able to trace the 
squared transmittion coefficient down to values of $10^{-20}-10^{-26}$
which implies that nearly the whole Brillouin zone can be studied.

Of course the presented method can be used in lattices with
dimension larger than one too. The analysis of the solutions however
will be much more tedious, thus we have demonstrated the
method using one-dimensional systems. The formulation of the
variational approach in the previous chapter can be also
useful in lattice dimensions higher than one.

We have used the results obtained with the new method in order
to test predictions of analytical considerations of the
spatial decay of a localized vibration. As a result we found
that nonlinear corrections of a linearization method used
to account for the exponential decay appear if a multiple
of the fundamental frequency of the localized solution
comes close to the phonon band. We have further shown how
these nonlinear corrections can be obtained analytically
and found very good agreement with the numerical results.
This part of the work was also carried out for one-dimensional
systems. It is possible to analyze higher dimensional lattices
with this method too. For that one has to consider lattice
Green's functions (see e.g. \cite{th91},\cite{dpw92}). These functions will
appear if one deals with the equations for the Fourier
components of a localized vibration on a given lattice
(the equations are essentially as in (\ref{9})).
Far away from the center of the NLE one can again linearize
these equations. The lattice Green's functions can be used then
in order to obtain analytical solutions. The $k$-dependent spatial
decay will be obtained, and nonlinear corrections can be
considered in full equivalence to the cases analyzed in the
present work.

Let us finally stress that the method and results obtained
in this work can be used in order to perform a dynamical stability
analysis of the NLE solutions as indicated in \cite{fw8}. The corresponding
eigenvalue problem can be formulated in the space of Fourier
coefficients. As an input of this stability analysis one needs
precise data on the Fourier coefficients of the periodic NLE solution,
which are provided by the presented method.
\\
\\
\\
\\
Acknowledgements
\\
\\
It is a pleasure to thank C. R. Willis, E. Olbrich and
K. Kladko for interesting and stimulating discussions,
and C. R. Willis and P. Fulde for a critical reading
of the manuscript.

\newpage

FIGURE CAPTIONS
\\
\\
\\
Fig.1
\\
\\
Numerical result for an XC NLE solution for model I (see text).
\\
Open squares - numerical data.
\\
Lines connect Fourier coefficients with same value of $k$.
\\
\\
\\
Fig.2
\\
\\
Same as in Fig.1 but for an XN NLE solution.
\\
\\
\\
Fig.3
\\
\\
Same as in Figs.1,2 but for an XC NLE solution of model II.
\\
\\
\\
Fig.4
\\
\\
Same as in Figs.1,2,3 but for an XC NLE solution of model III.
\\
\\
\\
Fig.5
\\
\\
Exponents of the spatial decay ${\rm ln}|\lambda|$ as a function
of $k$ for the NLE solution shown in Fig.3. \\
Open squares - numerical result. \\
Filled circles - linearization result.
\\
\\
\\
Fig.6
\\
\\
Same as in Fig.5 but for the NLE solution shown in Fig.4.
\\
\\
\\
Fig.7
\\
\\
$k$-dependence of the Fourier coefficients of the NLE solution
shown in Fig.1. Here the Fourier coefficients are renormalized
using the data for the central particle. The result
is plotted in a log-log plot. \\
Open squares - numerical data. \\
Lines - best eye fit of a power law (straight line in the plot)
for every lattice site $l$.
\\
\\
\\
Fig.8
\\
\\
The difference of the power law exponents
for the $k$-dependence of the Fourier coefficients for neighbouring
lattice sites as obtained from Fig.7. 

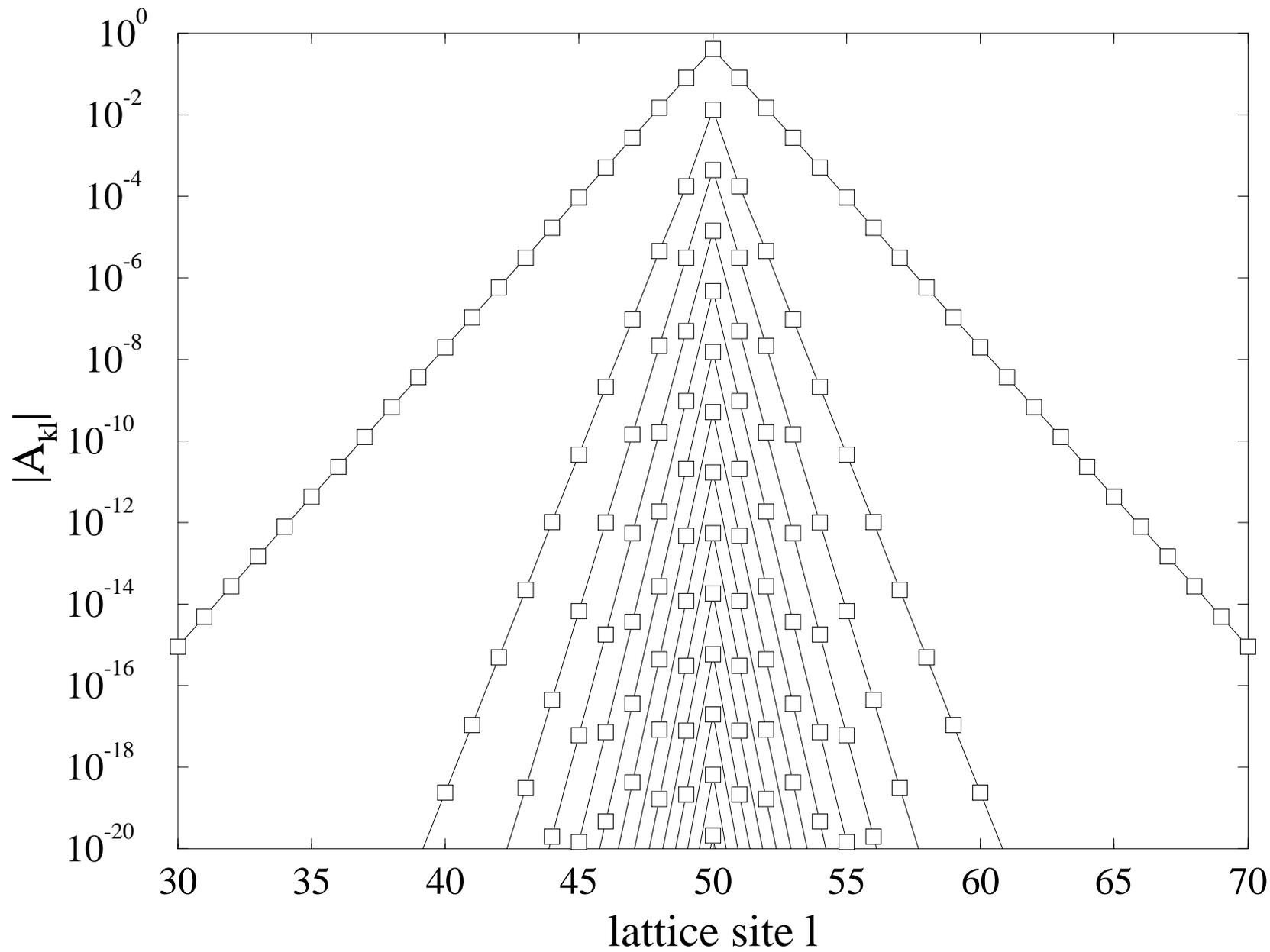
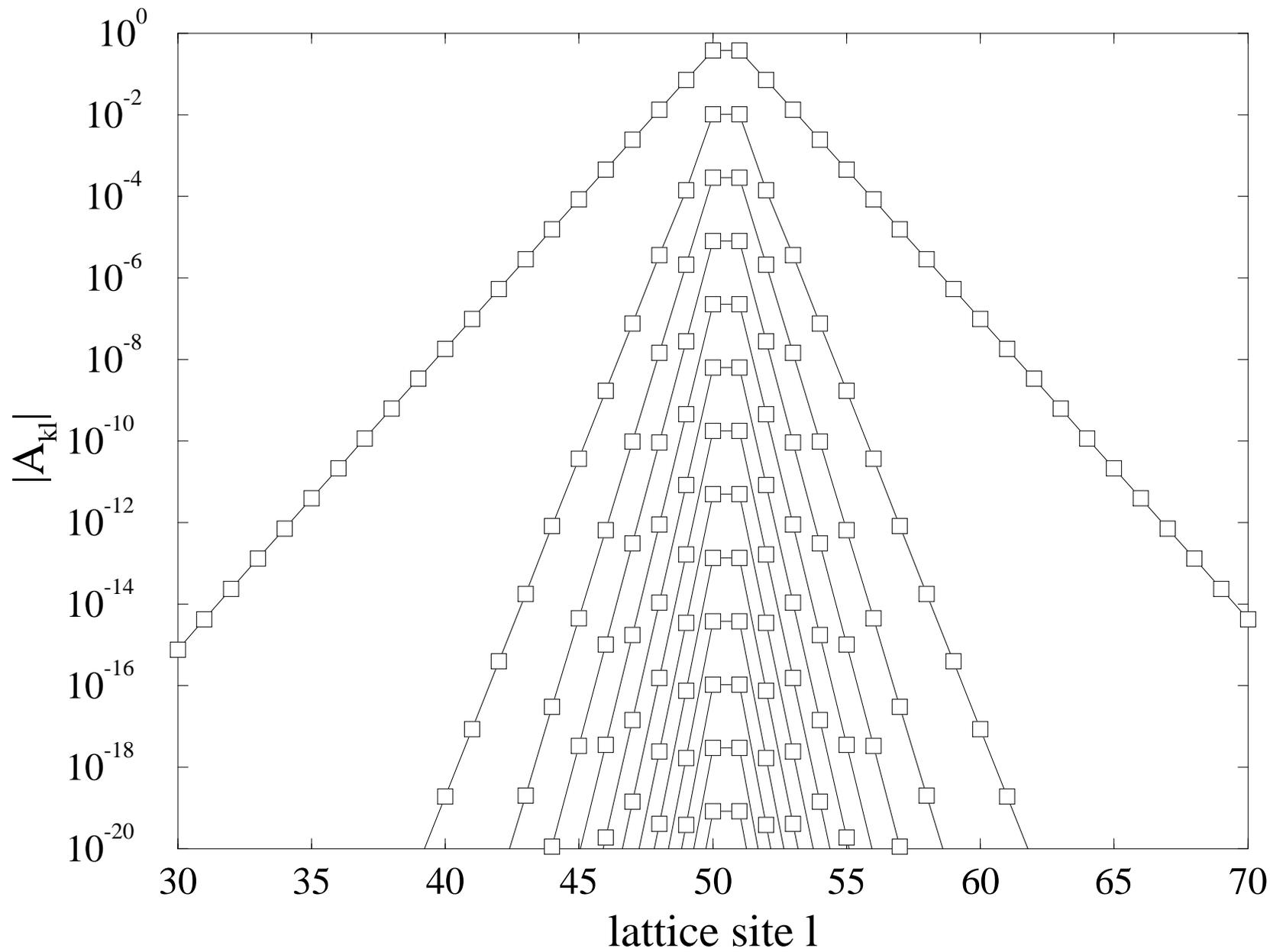
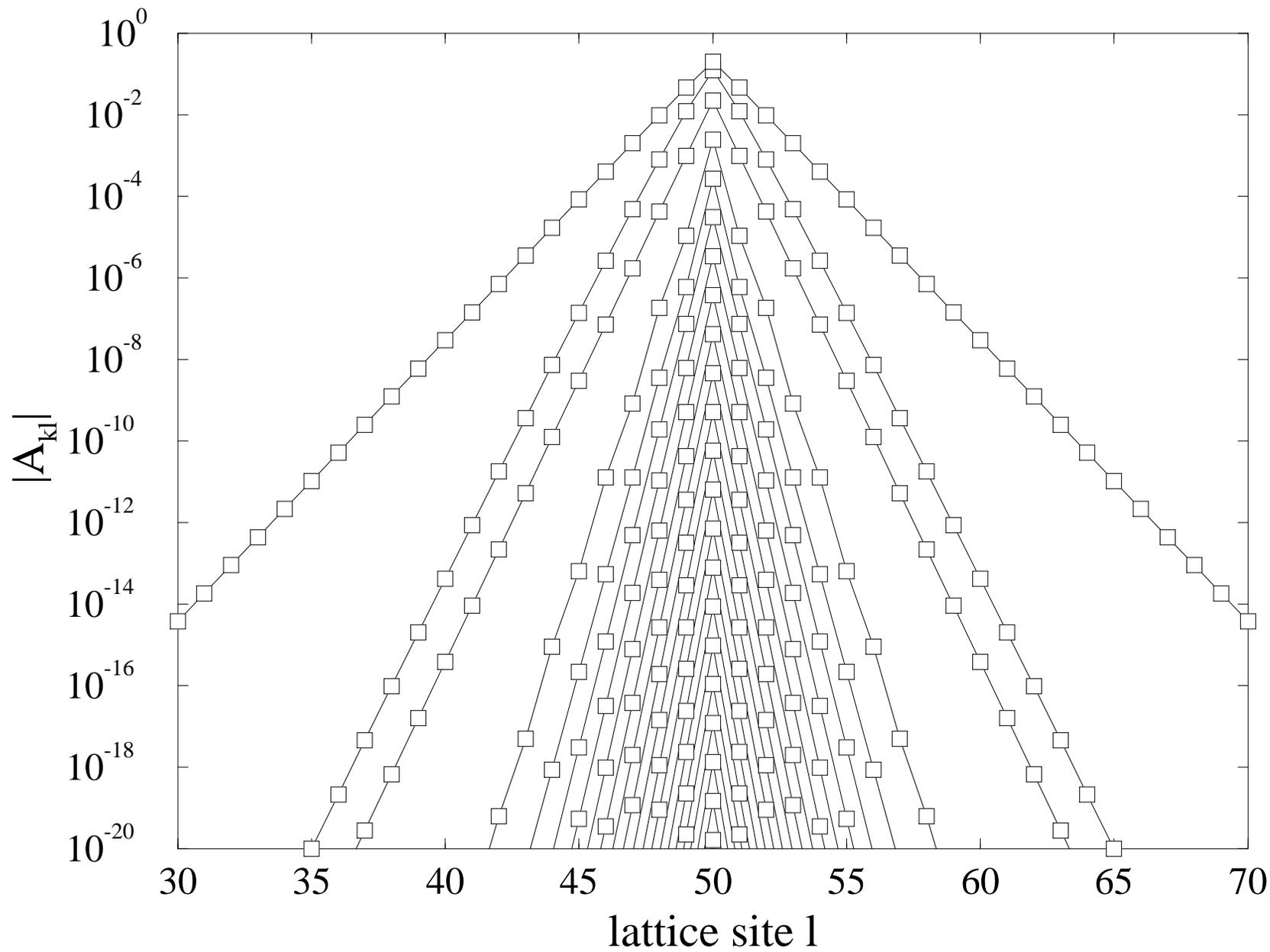
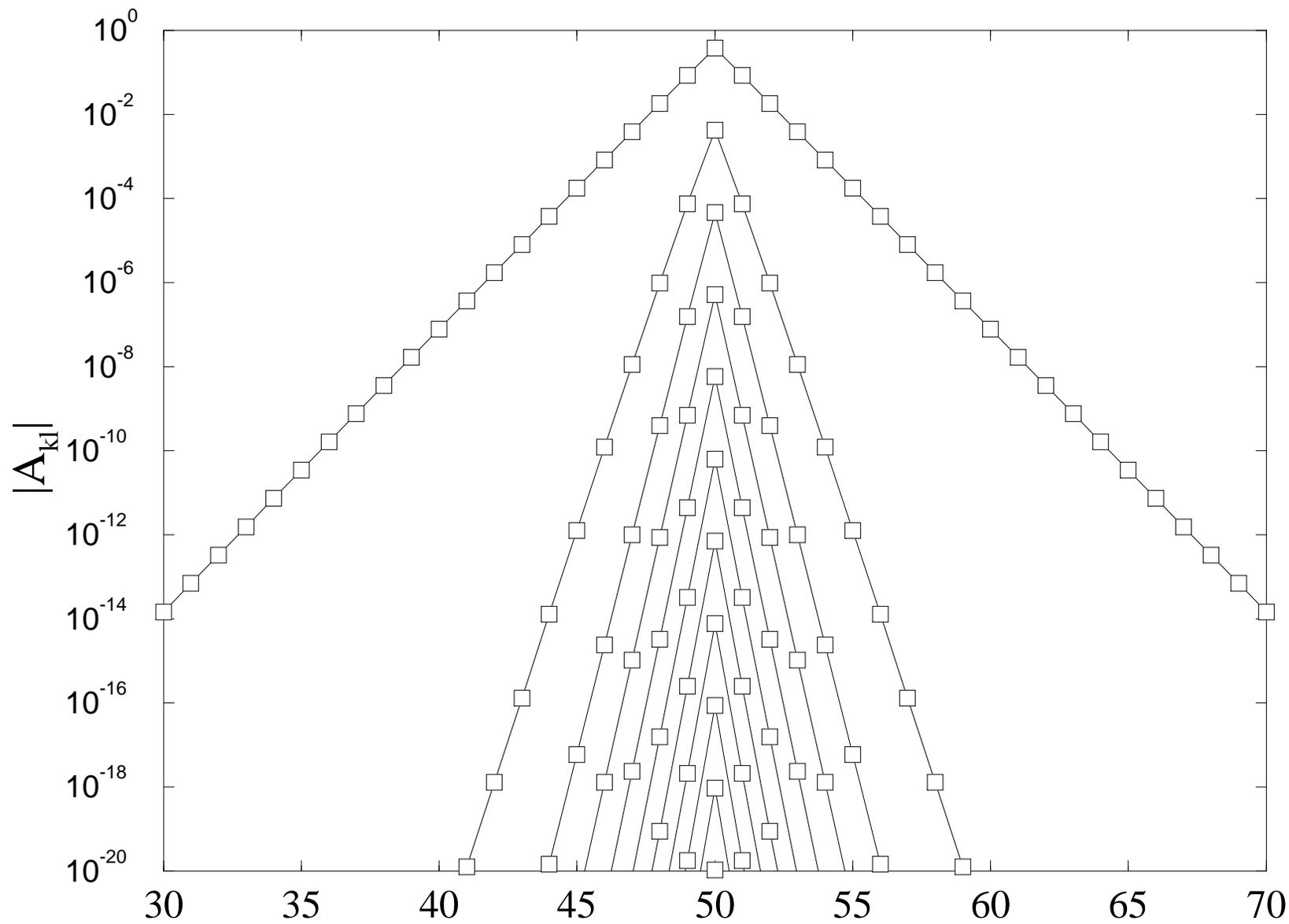
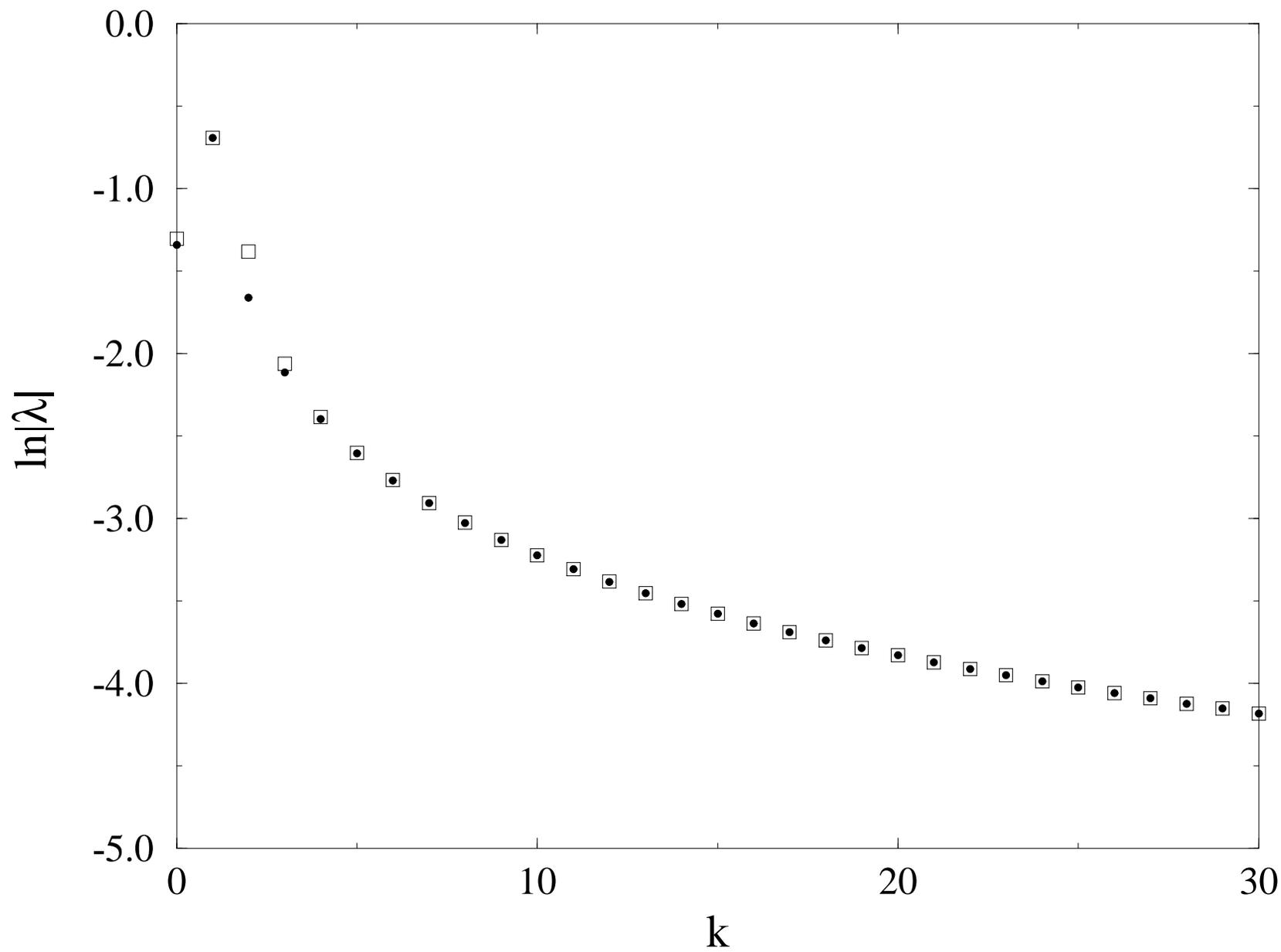
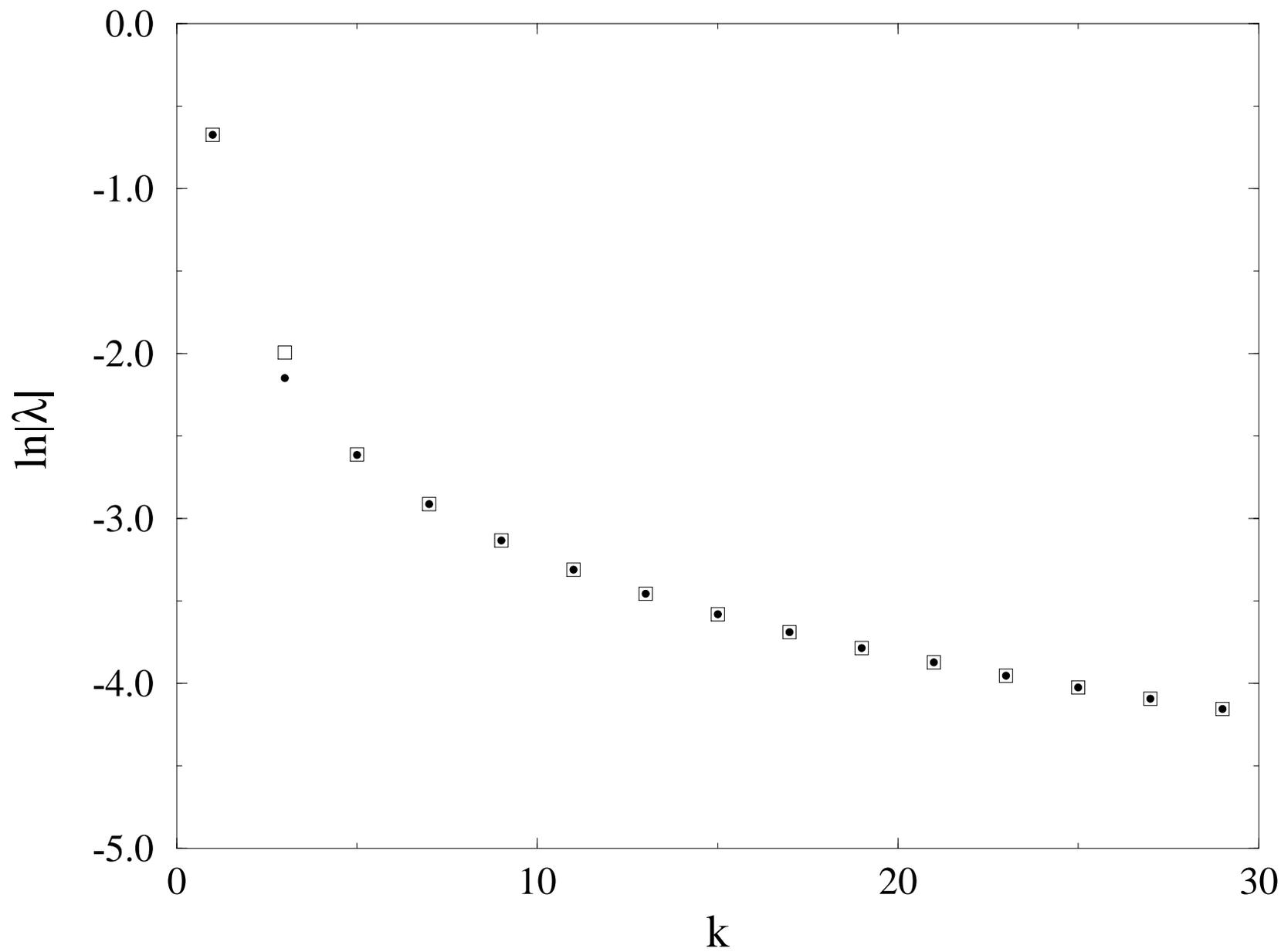
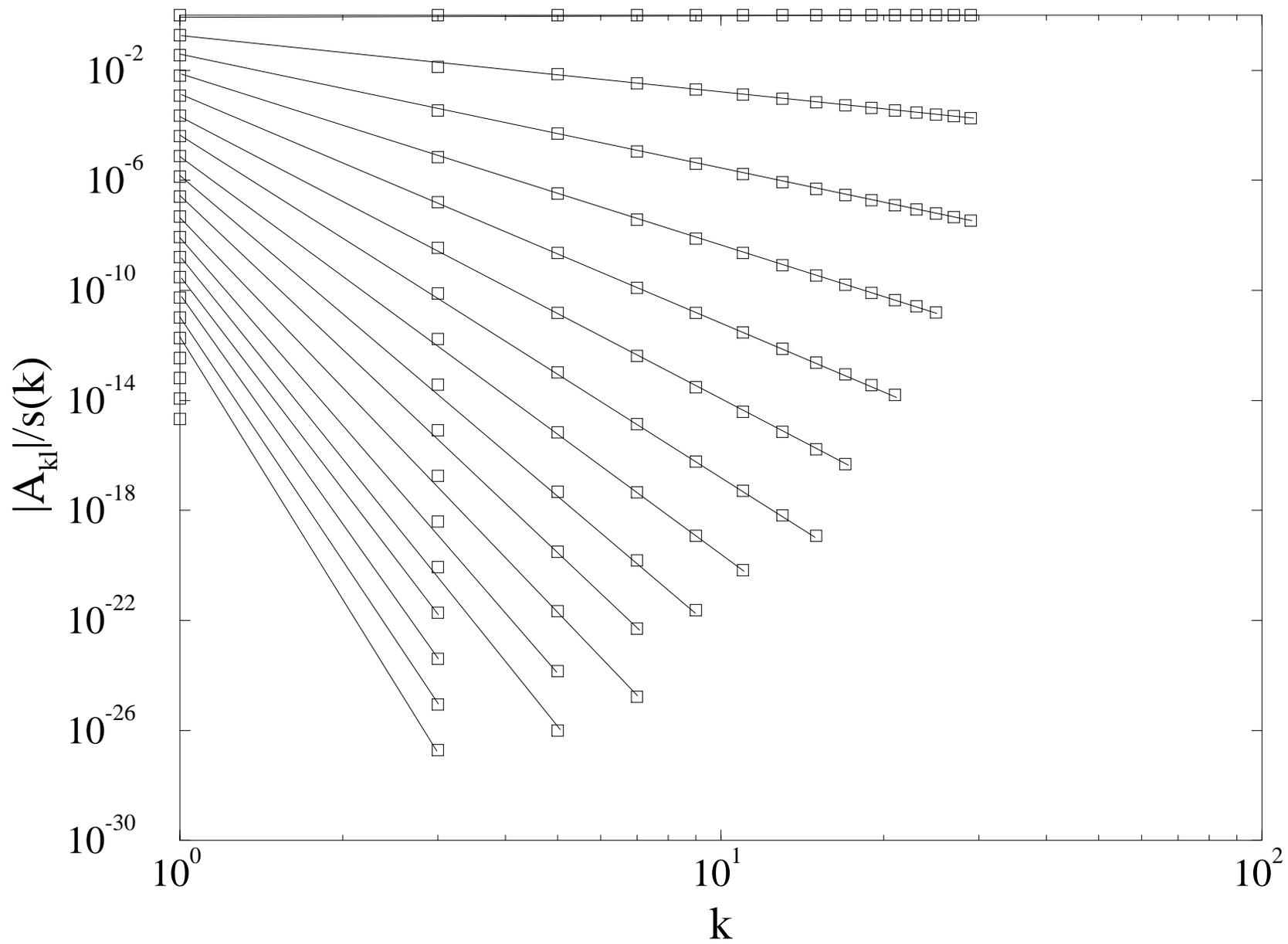
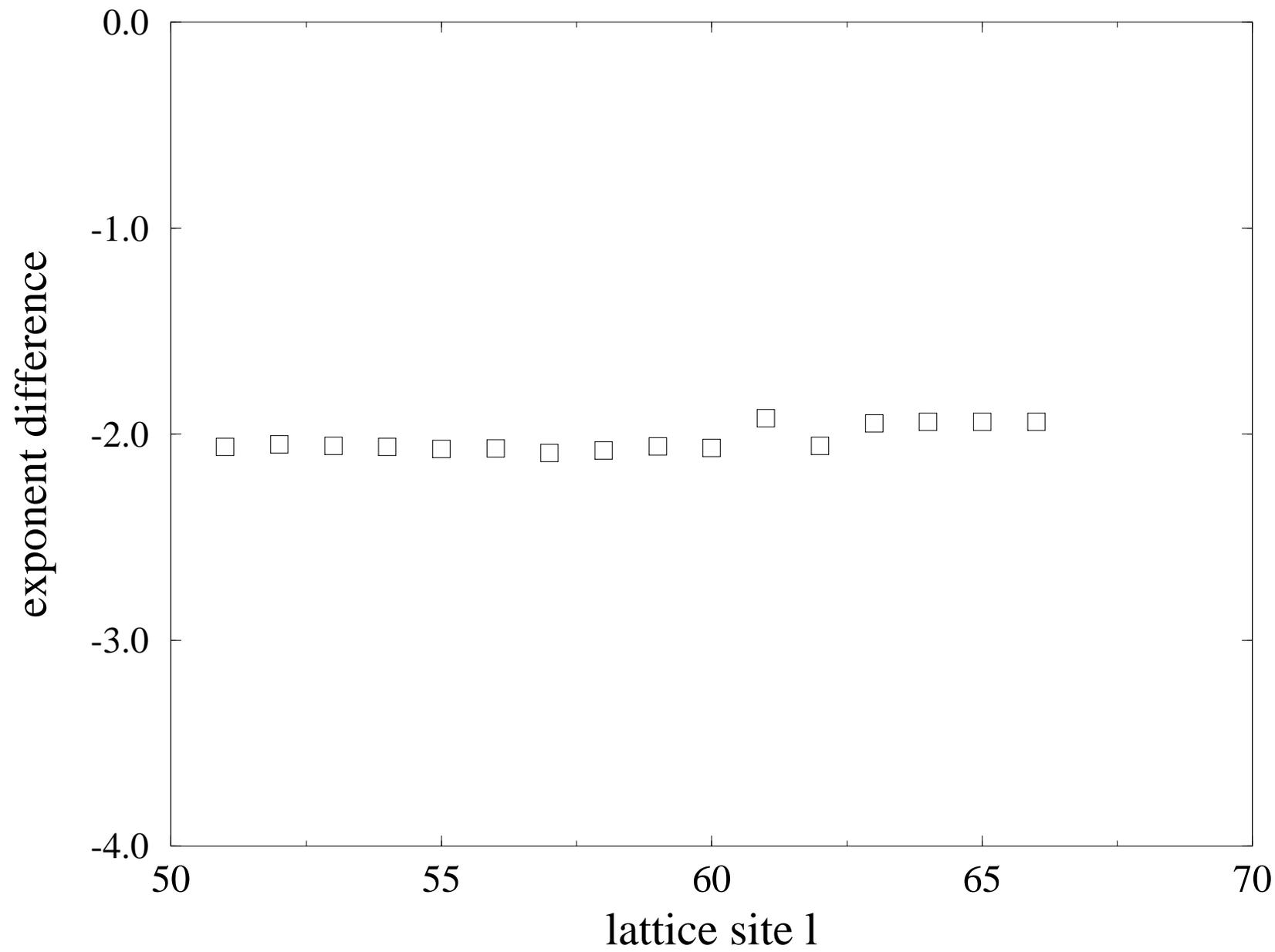
\end{document}